\documentclass{jpp}
\usepackage{graphicx}
\usepackage{epstopdf, epsfig}
\usepackage[utf8]{inputenc}
\usepackage[T1]{fontenc}
\usepackage{amsmath}
\usepackage{xcolor}

\shorttitle{Compton driven beam formation and magnetisation}
\shortauthor{B. Martinez, T. Grismayer and L.~O. Silva}

\title{Compton driven beam formation and magnetisation via plasma microinstabilities}

\author{Bertrand Martinez
  \corresp{\email{bertrand.martinez@tecnico.ulisboa.pt}},
  Thomas Grismayer
 \and Lu\'{i}s O. Silva}

\affiliation{GoLP/Instituto de Plasmas e Fusão Nuclear, Instituto Superior Técnico, Universidade de Lisboa, 1049-001 Lisbon, Portugal}

\begin{document}

\maketitle

\begin{abstract}
Compton scattering of gamma rays propagating in a pair plasma can drive the formation of a relativistic electron positron beam.
This process is scrutinised theoretically and numerically via particle-in-cell simulations.
In addition, we determine in which conditions the beam can prompt a beam-plasma instability and convert its kinetic energy into magnetic energy.
We argue that such conditions can be met at the photosphere radius of bright Gamma Ray Bursts.
\end{abstract}

\section{Introduction \label{sec:01}}

The interaction of gamma-rays with a pair plasma is a fundamental problem in astrophysics.
For instance, it is present in the context of Gamma-ray Bursts (GRBs).
The latter involve the explosion of a stellar mass object, which energy is expelled in the form of a relativistic ejecta.
As this ejecta propagates, its interaction with the ambient medium generates a strong emission which is detected by satellites as well as ground-based observatories.
Recent efforts \textcolor{black}{demonstrated emission over} a large range of frequencies including the radio, optical, x-ray and gamma-ray bands, up to TeV energies~\citep{NatAcciari2019}.

For most GRBs, the temporal evolution of the observations shows that their spectra consists of two parts, a prompt emission, which is followed by an afterglow.
The prompt emission of a GRB is defined as an initial emission phase in the keV-MeV band, lasting a few milliseconds to several minutes.
The afterglow is a counterpart of the prompt emission extending in the optical and radio ranges and its duration can span from a few hours to several days.
While it is acknowledged that a large part of the afterglow can be attributed to a forward external shock driven by the ejecta propagating in the ambient medium~\citep{APJMeszaros1997}, the origin of the prompt emission is not fully understood yet.
The main reason is that there remains unanswered questions related to the composition of the ejecta, how its energy is dissipated and how particles are accelerated.

All models agree that there is an ejecta formed by a compact object of radius $\sim 10^{7} \, \rm cm$.
This ejecta is a relativistic plasma in expansion.
Most of its energy comes from photons, but it also contains pairs, an unknown fraction of baryons and may even be magnetised.
At this early stage of expansion it is so dense that photons cannot escape.
At the photospheric radius $\sim 10^{12} \, \rm cm$, its density has decreased enough to enable a fraction of gamma-rays to propagate in the ambient medium~\citep{MNRASCavallo1978}.
One of the pending questions related to GRBs is how does the ejecta dissipates its energy above such a large distance of $\sim 10^{12} \, \rm cm$ ?
This question has been addressed in two ways, as detailed in this recent review on GRBs~\citep{PRKumar2015}.

Firstly, they brought forward models that describe the overall evolution of the ejecta.
The most used is the hot fireball model which assumes the energy of the ejecta is dissipated at the photosphere radius and in internal/external shocks~\citep{APJNarayan1992,APJRees1994,RMPPiran2005}.
However, it is worthwhile to mention that some GRB spectra are better explained with a model assuming a magnetized ejecta, where the energy is dissipated via current-driven instabilities~\citep{NJPLyutikov2006,APJZhang2010}.
The gamma-ray emission processes are manifold and are widely discussed in these models.
It is usually synchrotron~\citep{APJMeszaros1993,APJRees1994}, Inverse Compton~\citep{APJGruzinov2000} or inverse Compton up-scattering of synchrotron photons by high-energy electrons (synchrotron-self-Compton).
\textcolor{black}{
In the Klein-Nishina regime, the Compton process is also held responsible for a bulk plasma acceleration called Compton drag or radiative acceleration~\citep{APJMadau2000a,APJThompson2000}.
Our work here goes beyond these, by assuming that collective plasma processes can be triggered as the energetic photons propagate through the plasma.}
\textcolor{black}{On another hand}, some hadronic processes (Bethe-Heitler and photo-pion) are discussed as possible sources of high energy positrons, which then radiate gamma-rays via synchrotron.
\textcolor{black}{
Pair production processes are also proven to play a significant role into the development of key features of GRBs such as their high radiative efficiency~\citep{MNRASStern2003} or their flat spectrum from the infrared to the ultraviolet~\citep{APJBeloborodov2005}.
}

Secondly, researchers also performed numerical simulations of basic plasma processes in the extreme conditions of GRBs using Particle-In-cell (PIC) simulations.
Employing this framework enables to account for collective plasma dynamics, radiative processes and, more importantly, for any dynamical feedback between both.
The first example relevant in the context of GRB~\citep{APJMedvedev1999} is the simulation of relativistic collisionless shock formation by the Weibel instability~\citep{APJSilva2003} and the subsequent Fermi-type particle acceleration~\citep{APJSpitkovsky2008,APJMartins2009}.
The second example is particle acceleration in magnetic reconnection regions~\citep{MNRASMehlhaff2020} which is known to be of interest for GRBs~\citep{SSRUzdensky2011}.
The third one is electron acceleration in Compton driven plasma wakefields during the GRB interaction with its Circum Burst Medium (CBM)~\citep{APJFrederiksen2008, PRLGaudio2020}.

In his study, Frederiksen considers the interaction of a planckian photon distribution with a tenuous plasma~\citep{APJFrederiksen2008}.
The photon energy lies between $30\, \rm keV$ and $3 \, \rm MeV$ and he observes electron acceleration in a plasma wakefield.
According to the author, this wakefield is excited by an electrostatic force acting to restore the charge separation induced by Compton deflections.
Contrary to this work, our investigation focuses on a regime where the Compton cross section is beamed, above $10 \, \rm MeV$.
In addition, we consider a range where Compton scattering prevails over the Bethe-Heitler pair production ($\gamma e \rightarrow e^+e^- e$), for energies below $100 \, \rm MeV$.
Formulas for all these processes can be found in~\cite{APJLightman1982}.
For this gamma-ray range $10-100 \, \rm MeV$, one expects Compton scattering to deflect electrons mainly forward such that they form a relativistic beam~\citep{PRLGaudio2020}.

In light of these previous results, we investigate the interaction of gamma-rays (10-100 MeV) with a background pair plasma.
In section~\ref{sec:02}, with the support of a 1D theoretical model and Particle-In-Cell (PIC) simulations, we evidence how a gamma-ray beam (10-100 MeV) propagating in a pair plasma can drive the formation of a relativistic and dense electron positron beam via Compton scattering.
In section~\ref{sec:03}, we specify under which conditions the pair beam can trigger a beam-plasma instability, thus generating a small scale magnetic field which extent is discussed.
Section~\ref{sec:04} confirms the robustness of this process for various photon sources.
Finally, we argue in section~\ref{sec:05} that these conditions can be met \textcolor{black}{for radii $r$ such that $r \geq r_{\rm ph}$, with $r_{\rm ph}$ the photospheric radius of a GRB.}

\section{Pair beam formation \label{sec:02}}

\textcolor{black}{We first discuss a simplified physical picture}.
Let us consider a semi-infinite gamma-ray beam of density $n_{\omega 0}$ with a monoenergetic distribution.
We denote by $\epsilon = \hbar \omega / mc^2$ the photon energy normalized by the electron rest mass.
The photon beam propagates along the $x$ direction in an infinite cold pair plasma at rest, with density $n_{p0}$ associated to an angular frequency $\omega_p$.
In the frame of this work, we focus on the energy range $1 \ll \epsilon < 1/ \alpha_f = 137$\textcolor{black}{, where $\alpha_f$ denotes the fine structure constant}.
In this range, Compton scattering prevails over the two photon Breit-Wheeler process, and pair creation in the Coulomb field of an electron or positron~\citep{APJLightman1982}.

The formation process of the pair beam relies on the beaming of the Compton cross section for high-energy photons. 
Let us consider an electron at rest experiencing Compton deflection from a gamma-ray of energy $\epsilon$, with $d\sigma_{\rm kn}/d\Omega$ and $\sigma_{\rm kn}(\epsilon)$ the angular-differential and total Compton cross sections~\citep{NatKlein1928}, and $\theta$ the polar angle associated to $\Omega$, the photon angle after one scattering.
The Lorentz factor of the electron after the deflection $\gamma$ can be deduced from the energy and momentum balance as $\gamma / \epsilon = 1+1/\epsilon-1/\left[1+\epsilon(1-\cos \theta)\right]$. We introduce $\phi$ the angle between the deflected electron momentum and the incident photon direction and obtain $\mathrm{cotan} (\phi) = (1+\epsilon) \tan(\theta / 2)$.
Averaging those quantities over the Compton angular cross section, we obtain
for $\epsilon \gg 1$, $\langle \gamma \rangle_{\Omega} \simeq \epsilon$ and $\langle \sin^2 \phi \rangle_{\Omega} \simeq 4/\epsilon$~\citep{RMPBlumenthal1970}.
\textcolor{black}{This shows} that there is a simultaneous beaming of photons, electrons and positrons centered on the direction of the incident gamma-ray.
\textcolor{black}{We will discuss later the effects induced by a more realistic gamma-ray beam distribution.}

We assume the longitudinal momentum of the pair beam ($p_x$) can be approximated by its average over the Compton cross section $p_x  \simeq \sqrt{\langle p_{x}^2 \rangle_{\Omega}}$, and we define the transverse momentum spread induced by Compton scattering as $\Delta p_{\perp} = \sqrt{\langle p_{\perp}^2 \rangle_{\Omega}}$. For $\epsilon \gg 1$:
\begin{equation} 
    \frac{p_x}{mc} \simeq \epsilon \qquad \text{and} \qquad \frac{\Delta p_{\perp}}{mc} \simeq \sqrt{\frac{7 \epsilon}{6 \ln \epsilon}} \; \label{eq:01} .
\end{equation}
Equation~\eqref{eq:01} shows that the deflected electron energy can be as high as the incoming photon energy and that the transverse momentum spread of the pair beam is typically a few percent of its longitudinal momentum $\Delta p_{\perp}/p_x \simeq 5-20 \%$, for $ 1 \ll \epsilon < 1/\alpha_f $.

The density of the photons decreases as they are scattered at a frequency $ \tau_{\omega}^{-1} = 2 n_{p0} c \sigma_{\rm kn}$.
It thus follows that $dn_{\omega}/dt=-n_{\omega}/\tau_{\omega}$.
The solution reads $n_{\omega} / n_{\omega 0} = \exp(-t/\tau_{\omega})$.
In the high energy limit $\epsilon \gg 1$ one has $\sigma_{\rm kn}(\epsilon) \simeq \pi r_e^2 \ln(\epsilon)/\epsilon$, \textcolor{black}{with $r_e$ is the classical electron radius.
As a result, }the photon density can be approximated as constant $n_{\omega}/n_{\omega 0} \simeq 1$, where in fact $\omega_p \tau_{\omega} \propto r_e^{-3/2}n_{p0}^{-1/2} \epsilon / \ln(\epsilon) \gtrsim 10^{11}$ for any plasma density $n_{p0} \leq 10^{18} \, \rm cm^{-3}$, a conservative upper bound for astrophysical systems.

\textcolor{black}{
We now consider a more detailed model for the pair beam formation.
Let us denote $n_p(x,t)$ the background plasma density.
The electrons (and positrons) of the beam are the ones experiencing at least one Compton scattering.
Their density is denoted $n_b(x,t)$ and they have a velocity $v_b$.
At the initial time $t=0$, we assume the photons are located in the $x<0$ half space and propagate toward the background plasma at rest in the other half-space $x>0$.
The evolution of the background plasma density (beam density) is given by the continuity equation with a source term accounting for a depletion (loading) at the Compton frequency $ \nu=\sigma_{\rm kn} c n_{\omega} $
\begin{align}
  \partial_t n_p + \nabla .(n_p \vec{v}_p) & = - \sigma_{\rm kn} c n_\omega n_p  \nonumber \\
  \partial_t n_b + \nabla .(n_b \vec{v}_b) & = \sigma_{\rm kn} c n_\omega n_p  \nonumber
\end{align}
We introduce the Heaviside function $H$, along with the variables $\xi,\tau$ defined as $\xi=ct-x$ and $\tau=t$.
The background plasma is assumed to be at rest during the interaction ($\vec{v}_p=0$), and the beam to have a constant velocity $\vec{v}_b=v_x \vec{x}=c\beta_x\vec{x}$.
The evolution of the background plasma and beam densities simplifies to
\begin{align}
  \left[c\partial_\xi+\partial_\tau\right] n_p(\xi,\tau) & = - \nu n_p(\xi,\tau) \left[H(\xi)-H(\xi - ct)\right] \label{eq:n_back_equ} \\
  \left[(c-v_x)\partial_\xi+\partial_\tau\right] n_b(\xi,\tau) & = \nu n_p(\xi,\tau) \left[H(\xi)-H(\xi - ct)\right]  \label{eq:n_beam_equ} 
\end{align}
We first solved Eq.~\eqref{eq:n_back_equ} by assuming the solution is of the form $f(\xi)g(\tau)$.
As far as Eq.~\eqref{eq:n_beam_equ} is concerned, its solution was derived using the method of Laplace transform.
The solutions are obtained for all $\xi \leq c\tau$ as
\begin{align}
  n_p(\xi,\tau) = n_{p0}e^{ - \frac{\nu \xi}{c}} & \left[H(\xi)-H(\xi - c\tau)\right] \label{eq:n_back_sol} \\
  n_b(\xi,\tau) = \frac{n_{p0}}{1-\beta_x}\Big[ & \left( 1 -e^{-\frac{\nu \xi}{c}} \right)H(\xi) \label{eq:n_beam_sol} \\ 
  & + \left( e^{-\frac{\nu}{v_x}\left(\xi-c\tau(1-\beta_x)\right)}-1 \right)H(\xi-c\tau(1-\beta_x))  \Big] \nonumber
\end{align}
The normalized beam velocity $\beta_x$ can be estimated as $ \beta_x \simeq  \sqrt{ \langle \beta_x^2 \rangle_{\Omega}}$.
In the high energy limit $\epsilon \gg 1$, $\beta_x \simeq 1 - 1/ (2\epsilon\ln \epsilon)$.
For large times $ t \gg 1 / \left[\nu (1-\beta_x)\right]$, Eq.~\eqref{eq:n_beam_sol} implies that the beam cannot exceed the maximum density $ n_{p0}/(1-\beta_x)\simeq 2 n_{p0} \epsilon \log \epsilon$.
Since we expect collective processes for such time scales, we underline that this limit is unlikely to be reached and should be considered as an upper bound of the maximum achievable density.
This maximum density could however be achieved in cases where collective effects are damped on large distances, for example either with a high magnetic field oriented in the flow direction or a large background plasma temperature.
}

\textcolor{black}{
In order to evidence some of the properties of this solution, we recast it with $(x,t)$ variables and in the limit $\nu t\ll 1$ and $\beta_x\simeq 1$ and therefore get for all $ 0 \leq x \leq ct $
\begin{align}
  n_p(x,t) & = n_{p0} \left(1+\nu x/c - \nu t\right) \label{eq:n_back_sol_sim} \\
  n_b(x,t) & = n_{p0} \nu x / c \label{eq:n_beam_sol_sim}
\end{align}
The result in Eq.~\eqref{eq:n_back_sol_sim} illustrates that the background plasma density evolves linearly with the position $x=0\rightarrow ct$, increasing from $n_{p0}(1-\nu t)$ to its maximum value $n_{p0}$.
Eq.~\eqref{eq:n_beam_sol_sim} evidences that the beam density is expected to grow linearly with the propagation distance, starting from $0$ and increasing to its maximum value $n_{p0}\nu t$ at the position of the photon front $x=ct$.
}

We stress this model is valid whatever the photon and plasma densities.
It remains applicable as long as collective plasma effects do not play a significant role.
In terms of photon energies, it is strictly limited to the range $1 \ll \epsilon < 1/\alpha_f = 137$.
However, we will discuss in section~\ref{sec:04} why it can be extended to a larger range of $20 \lesssim \epsilon \lesssim 200$.
\textcolor{black}{In order to account for photon in the energy range $1 \lesssim \epsilon \lesssim 20$, one would need to account for the much larger energy spread induced by the Compton cross-section and also to include the two photon Breit-Wheeler pair creation.}

\begin{figure}
\includegraphics[width=\textwidth]{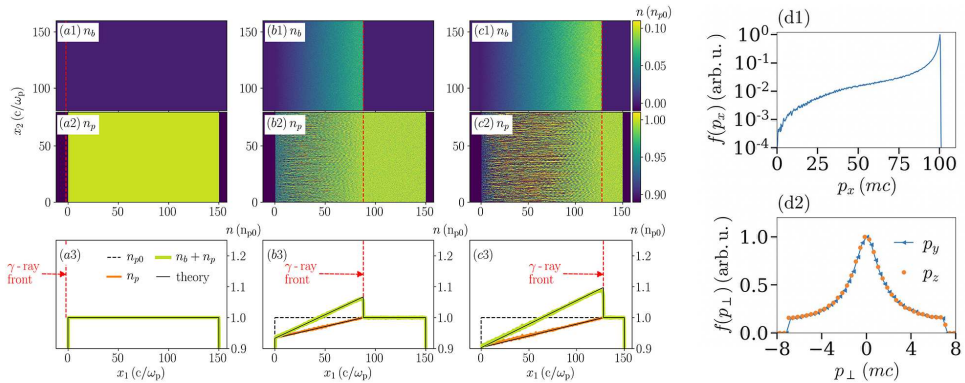}
\caption{
\label{fig:01}
(a1-b1-c1) illustrate the 2D electron \textcolor{black}{beam} density profile at three instants (a1) $\omega_p t = 0$ (b1) $\omega_p t = 90$ (c1) $\omega_p t = 130$.
\textcolor{black}{
(a2-b2-c2) display the 2D electron \textcolor{black}{background} density profile for the same three instants.}
(a3-b3-c3) are 1D projections along propagation direction $x$ for the same instants.
The orange curve is the background plasma density $n_p$, the green thick curve is the total density $n_p+n_b$ and the black curves are the theoretical estimates from Eqs.~\eqref{eq:n_back_sol}-\eqref{eq:n_beam_sol}.
The gamma-ray beam propagates from left to right and its front is marked by the dashed red line.
(d1-d2) $f(p_x)$ and $f(p_{\perp})$ distributions of electrons scattered by Compton at the beam front for time $\omega_p t=130$.
For the plasma density $n_{p0}=1 \, \rm cm^{-3}$, one has $c/\omega_p = 5.3\times 10^5 \, \rm cm$.
}
\end{figure}

We have reported how a pair beam can be created when gamma-rays propagate in a background pair plasma via Compton scattering.
This process is corroborated with theory and is now confronted to simulation results.
We have run 2D Particle-In-Cell (PIC) simulations with \textsc{Osiris}~\citep{PrFonseca2002}.
It was recently enriched with a Compton scattering module~\citep{JPPGaudio2020},
similarly to earlier work~\citep{PoPHaugbolle2013}.
It proceeds through two key steps: first a random pairing of macro particles at every time step and in every cells and second a Monte Carlo sampling of the angle-resolved Klein-Nishina cross section~\citep{NatKlein1928}. For every binary collisions treated, the cross section is evaluated in the electron rest-frame, an energy and momentum balance is ensured and a splitting technique is employed to handle the varying weight of macro particles.
This implementation relies on macro particle pairing and therefore stands out from other works~\citep{AALevinson2018} where the Compton cross section is evaluated from the intensity of the ambient radiation field.

We detail the relativistic pair beam formation on an illustrative simulation in Fig.~\ref{fig:01}.
We consider a cold pair plasma ($1 \, \rm eV$) with a uniform density $n_{p0}=1 \, \rm cm^{-3}$, as displayed in Figs.~\ref{fig:01}(a1-a3).
In this pair plasma, we propagate a monoenergetic ($\epsilon = 100$) gamma-ray beam along the $x$ direction from left to right. It presents a uniform density $n_{\omega 0} = 10^{17} \, \rm cm^{-3}$ with a front indicated by the \textcolor{black}{red} dashed line in \textcolor{black}{Figs.~\ref{fig:01}(a3,b3,c3)}.
Periodic boundary conditions are assumed in the transverse direction $y$ for all species and fields \textcolor{black}{and open boundary conditions are employed in the longitudinal $x$ direction}.
The domain extends on $(160 \, c/\omega_p)^2$ with $2000^2$ cells.
The cell dimensions are $\delta x = \delta y = 0.08 \, c/\omega_p $ and the time step is $\delta t = \delta x /2$.
We initialize $80/80/160$ particles per cell for electrons, positrons and photons.

The process of pair beam formation is illustrated in Fig.~\ref{fig:01}.
Figs.~\ref{fig:01}(a1-b1-c1) exhibit 2D profiles of the electron \textcolor{black}{beam} density $n_b$ at three instants $\omega_p t = 0$, $90$ and $130$.
\textcolor{black}{
They correspond to electrons which experienced at least one Compton scattering.
Figs.~\ref{fig:01}(a2-b2-c2) represent 2D profiles of the electron \textcolor{black}{background} density $n_p$ for the same three instants $\omega_p t = 0$, $90$ and $130$.}
Figs.~\ref{fig:01}(a3-b3-c3) are averages of theses 2D densities along $y$ the transverse direction.
\textcolor{black}{
To clarify the plot, only the total density $n_b+n_p$ (green thick curve) and background density $n_p$ (orange thin curve) are represented.}
Based on these figures, we can observe the formation and propagation of an electron beam (positrons are superimposed by symmetry).
We found excellent agreement of the simulations with the theoretical estimates (black lines) of Eqs.~\eqref{eq:n_back_sol}-\eqref{eq:n_beam_sol}.
\textcolor{black}{ 
The simulation times considered here ($\omega_p t \leq 130$) lie within the limit $\nu t \ll 1$ where we derived Eqs.~\eqref{eq:n_back_sol_sim}-\eqref{eq:n_beam_sol_sim}.
For such early times, the beam and background densities have profiles increasing linearly with the propagation distance, which explains the 
triangular shape of the total density profile in Figs.~\ref{fig:01}(b3-c3).
}
The momenta distributions $f(p_x)$ and $f(p_{\perp})$ of deflected electrons are exemplified in Figs.~\ref{fig:01}(d1-d2) at time $\omega_p t = 130$.
The distribution $f(p_x)$ exposes a peak at $p_x/mc \simeq 100 = \epsilon$, exactly as our theoretical estimate in Eq.~\eqref{eq:01}.
The transverse momentum profile $f(p_{\perp})$ includes both directions $p_y$ and $p_z$.
The two distributions are centered and characterized by a $\simeq 3 \, mc$ standard deviation, in agreement with Eq.~\eqref{eq:01}.
This is conducted in the range of validity of the theoretical model with periodic conditions in the transverse direction, and does not evidence collective plasma processes with a significant impact on the beam formation.

Another set of 2D simulations was run to assert the relevance of this process of beam formation.
Our goal is to check the range of validity of the scaling inferred for its peak density in Eq.~\eqref{eq:n_beam_sol}.
We consider two sets of simulations.
The first set is characterised by a pair plasma of density $n_{p0}=10^{18} \, \rm cm^{-3}$, with photon densities $n_{\omega 0}= 10^{25}$-$10^{27}\, \rm cm^{-3}$.
The second presents a pair plasma density of $n_{p0}=1 \, \rm cm^{-3}$, with photon densities $n_{\omega 0} = 10^{16}$-$10^{18}\, \rm cm^{-3}$.
Connections of these prameters to laboratory or astrophysical parameters will be discussed in section~\ref{sec:05}.
Given the symmetry of our problem, we reduce the domain size to $(24 \, c/\omega_p)^2$ and follow the photon beam in a moving window \textcolor{black}{with a velocity equal to $c$}.
The domain has $600^2$ cells with dimensions $\delta x = \delta y = 0.04 \, c/\omega_p$ and a time step $\delta t = \delta x /2$.

\begin{figure}
\centering
\includegraphics[width=0.60\textwidth]{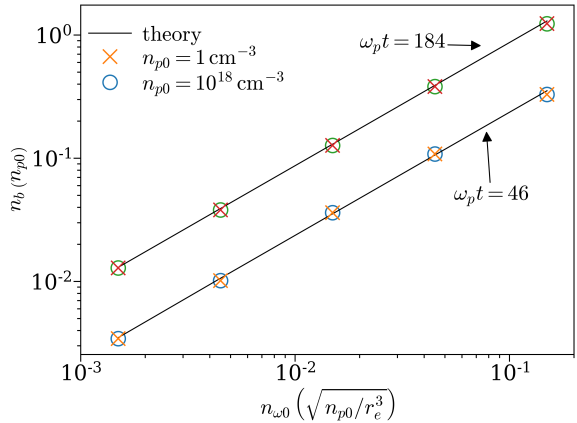}
\caption{
\label{fig:02}
Pair beam density $n_b/n_{p0}$ versus the photon density $n_{\omega 0}$ in unit of $\sqrt{n_{p0}/r_e^3}$ for plasma densities $n_{p0} = 1 \, \rm cm^{-3}$ (X) and $n_{p0} = 10^{18} \, \rm cm^{-3}$ (O) at two times in the simulations.
\textcolor{black}{$\omega_p $ is the plasma angular frequency.}
}
\end{figure}

Figure~\ref{fig:02} represents the pair beam density as a function of the incident photon density for two instants, \textcolor{black}{$ \omega_p t = 46$ and $184$.
The normalization by the quantity $(n_{p0}/r_e^3)^{1/2}$ comes from a factor $\nu/\omega_p$ in Eq.~\eqref{eq:n_beam_sol_sim}.}
It enables to reveal a general scaling of the normalized pair beam density $n_b/n_{p0}$, independent of the initial plasma density $n_{p0}$.
Despite such a wide gap of initial plasma densities considered in Fig.~\ref{fig:02}, simulations evidence that the normalized pair beam density at a given time $t$ remains exactly the same as long as $n_{\omega 0}(r_e^3/n_{p0})^{1/2}$ (or $\nu/\omega_p$) is constant.
This scaling matches exactly with the prediction of Eq.~\eqref{eq:n_beam_sol}, illustrated as a black line in Fig.~\ref{fig:02}.

We have generalised the validity of the results obtained in Fig.~\ref{fig:02}, to account for baryon loading in the background plasma.
Firstly, we ran all the simulations presented before with a fraction $f$ of proton density up to $f=0.1 n_{p0}$ in the pair plasma.
\textcolor{black}{
As the initial electron and positron plasma densities are not equal, the beam driven by Compton now presents a non-uniform charge and current densities.
This initiates the start of the two-stream instability.
The growth rate of such instability is much lower than the oblique instability~\citep{PREBret2005} by a factor $\gamma_b^{2/3}$, with $\gamma_b$ the Lorentz factor of the beam.
As a consequence, the exponential growth of the oblique modes we report in the next section still takes place, but is initiated in a slightly perturbed plasma.
}
Secondly, we ran one 3D simulation, confirming our previous findings.
We performed it for a plasma density of $n_{p0}=1 \, \rm cm^{-3}$ and photon density of $n_{\omega 0} =10^{18}\, \rm cm^{-3}$.
The domain size is $(24 \, c/\omega_p)^3$ and the photon beam is followed in a moving window \textcolor{black}{with a velocity equal to $c$}.
The domain has $600^3$ cells of dimensions $\delta x = \delta y = \delta z = 0.04 \, c/\omega_p$ and the time step is $\delta t = \delta x /2$.
We initialize $2/2/4$ particles per cell for electrons, positrons and photons.
The transverse boundary conditions for particles and fields are all periodic.
For this 3D simulation \textcolor{black}{(not shown here)}, the evolution of the peak beam density is the same as for the 2D simulation with the same parameters.
While we will discuss this in detail in section~\ref{sec:03}, we add that the growth rate of the beam plasma instability in this 3D simulation is also the same as in the corresponding 2D simulation.

\section{Onset of a beam-plasma instability \label{sec:03}}

The normalized pair beam density $n_b/n_{p0}$ increases with time and is expected to trigger \textcolor{black}{an electromagnetic beam-plasma instability when $n_b/n_{p0} \lesssim 1$}.
Two types of modes can compete and lead to an exponential growth of magnetic fields~\citep{PRLBret2005}.
Firstly, \textcolor{black}{the modes associated to} the Current Filamentation Instability (CFI), with growth rate $\omega_p^{-1}\Gamma_{\rm CFI} = (n_b /n_{p0}\gamma_b)^{1/2}$.
Secondly, \textcolor{black}{the modes corresponding to} the OBlique filamentation Instability (OBI), with growth rate $\omega_p^{-1}\Gamma_{\rm OBI} = \sqrt{3}(n_b /n_{p0}\gamma_b)^{1/3} /2^{4/3}$.
We can predict that our prevailing modes will be oblique since we consider Lorentz factors $ 1 \ll \gamma_b \lesssim 1/ \alpha_f$ and densities $n_b /n_{p0} \leq 1$.
Above some critical angle in $\mathbf{k}$ space, such modes are expected to be damped like CFI modes due to thermal effects~\citep{PRLBret2005}.
The physical reason is that a transverse thermal expansion of filaments competes with the magnetic pinching force maintaining them~\citep{PoPSilva2002}.
The latter theoretical work provides a density threshold above which the instability can still set off despite this initial spread.
It reads $\alpha_{\rm th} = n_b/n_{p0} \simeq \gamma_b (p_{\perp}/mc\gamma_b)^2$.
Using the estimates from Eq.~\eqref{eq:01}, we get $n_b/n_{p0} \simeq 7 / (6 \ln \epsilon)$ and
\begin{equation}
  \omega_p^{-1}\Gamma \simeq (7 \sqrt{3}/16)^{1/3} [\epsilon \log(\epsilon)]^{-1/3} \; .
  \label{eq:04}
\end{equation}
\textcolor{black}{
The minimum density required to trigger the instability is reached after a propagation time of $ 1 / \left( \nu \ln \epsilon \right) $.
Using Eq.~\eqref{eq:n_beam_sol}, this implies the photons should propagate through a plasma of length greater than $ r_{\rm min} \geq c / \left( \nu \ln \epsilon \right) $.
This condition can be recasted as
\begin{equation}
  r_{\rm min} \, [\mathrm{cm}] \geq 1.6\times 10^{25} \, n_{\omega}^{-1} \, [\mathrm{cm^{-3}}]  \; .
  \label{eq:r_min}
\end{equation}
The beam density threshold $\alpha_{\rm th}=6/(7\ln\epsilon)$ is independent from the initial plasma density, and so is the plasma length $r_{\rm min}$ given in Eq.~\eqref{eq:r_min}.}
It is worthwhile to stress that provided the photon density and the background plasma length satifies the condition in Eq.~\eqref{eq:r_min}, the instability will take place.
\textcolor{black}{
As far as the effect of the background plasma temperature is concerned, we can rely on previous investigations~\citep{PREBret2005}.
For the non relativistic background plasma temperature considered here ($1 \, \rm eV$), we do not expect any mitigation effect on the onset of CFI-like modes.
}

In addition, it seems meaningful to evaluate the energy conversion efficiency from the incident photons to magnetic fields.
We define  $\eta = \mathcal{E}_{B_z}/\mathcal{E}_{\omega 0}$, the ratio between the magnetic energy $\mathcal{E}_{B_z} \simeq \int c^2 B_z^{2}/8\pi \, d\mathbf{x}$ and the incident photon energy.
For a semi-infinite beam (and 1D), the ratio $\eta$ is also the ratio of the energy densities.
For the sake of simplicity, we estimate an upper bound for the magnetic field energy $\mathcal{E}_{B_z} \lesssim \int c^2 |B_z|^{2}/8\pi \, d\mathbf{x}$.
Using Parseval theorem, it can be expressed as $\mathcal{E}_{B_z} \lesssim \int c^2 |B_z|^{2}/8\pi \, d\mathbf{k}$.
The value of the saturated magnetic field can be approximated by equating the bounce frequency of trapped particles to the growth rate of the instability~\citep{PoPYang1994}.
This maximum is $eB_{z}(k)/m\omega_p \simeq (\gamma_b^2/p_x) (\omega_p^{-1} \Gamma)^2 \omega_p /kc$.
The energy conversion efficiency is then deduced by introducing the incident photon energy density $\epsilon n_{\omega 0}$
\begin{equation}
    \eta \lesssim \frac{1}{4 \pi} \left( \frac{7\sqrt{3}}{16} \right)^{4/3} \frac{n_{p0}}{n_{\omega 0}} \frac{1}{\left(\epsilon \ln^4 (\epsilon) \right)^{1/3}} \; .
    \label{eq:06}
\end{equation}
It is important to underline this inequality only expresses an upper bound for the conversion efficiency.
\textcolor{black}{
For any given plasma of density $n_{p0}$, Eq.~\eqref{eq:06} predicts that $\eta$ can be maximized for smaller incident photon densities $n_{\omega 0}$.
However, decreasing the photon density limits the growth of the beam density through Compton scattering and therefore increases the propagation distance required to trigger the instability, as evidenced in Eq.~\eqref{eq:r_min}.}
As a consequence of this low conversion efficiency $\eta$, the generation of magnetic fields can in principle take place over long distances.
For a perfectly collimated photon source emerging from the photosphere ($\simeq 10^{12}  \, \rm cm$), this length can be as high as the Compton mean free path $L_{\omega}\, [\mathrm{cm}] =c\tau_{\omega}  \simeq 10^{25} n_{p0}^{-1} \,\rm [cm^{-3}]$.
In fact, the typical opening angle of the GRB ejecta is in the range $\theta \leq 20^{\circ}$, as estimated in ~\cite{APJRacusin2009}.
\textcolor{black}{
With this more realistic assumption, the photon density and the beam density are expected to decrease with the propagation distance as $1/r^2$ due to this transverse dilution effect.
We estimate under which condition this depletion of the beam may remain negligible compared to  the loading via Compton scattering as
\begin{equation} \label{eq:r_max}
  \frac{\nu}{c} (r-r_{ph}) \left(\frac{r_{ph}}{r}\right)^2 \geq 1 \quad \rightarrow \quad r \leq \sigma_{\rm kn} n_{\omega 0} r_{ph}^2 = r_{\rm max}
\end{equation}
The inequality~\eqref{eq:r_max} states that the density increase due to the Compton loading between $r_{ph}$ and $r$ is $\nu (r-r_{ph})/c$ and remains larger than the density decrease between $r_{\rm ph}$ and $r$ due to the transverse dilution, which is $\left(r_{ph}/r\right)^2$.
The inequality~\eqref{eq:r_max} provides a maximum  distance, denoted $r_{\rm max}$, over which the instability can be sustained.
}

Given a gamma-ray density between \textcolor{black}{$10^{14}\, \rm cm^{-3}$} and $10^{22}\, \rm cm^{-3}$ that we inferred at the exit of the photosphere of GRBs (see section~\ref{sec:05}), the condition~\eqref{eq:r_max} is fulfilled for distances ranging from a few photospheric radii \textcolor{black}{$ \sim 10^{12} \, \rm cm$} up to larger values of \textcolor{black}{$ 10^{20} \, \rm cm$}.
\textcolor{black}{
We did not verified this estimate in simulations due to the computational cost for very long propagation lengths $L \, [\mathrm{cm}] \geq 10^{8} n_{p0}^{-1/2} \, \rm[cm^{-3}]$.}

\section{Generalisation for various photon distributions \label{sec:04}}

In the previous section, we have considered idealized photon distributions.
We now examine more realistic photon distribution functions.

\begin{figure}
\centering
\includegraphics[width=0.60\textwidth]{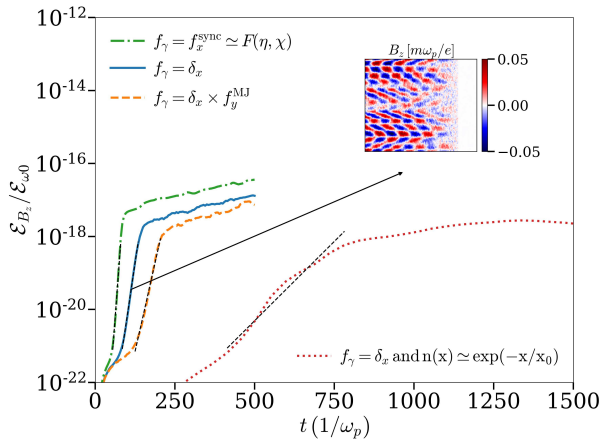}
\caption{
\label{fig:03}
Magnetic field energy $\mathcal{E}_{B_z}$, normalized by the incident photon energy $\mathcal{E}_{\omega 0}$.
\textcolor{black}{We denoted $\eta=\mathcal{E}_{B_z}/\mathcal{E}_{\omega 0}$ and derived an upper bound in Eq.~\eqref{eq:06}.}
Simulation for a monoenergetic photon distribution (blue solid curve),  a Maxwell-J\"{u}ttner transverse profile (orange dashed curve),  a front density gradient (red dotted curve), and a synchrotron distribution (green dash-dotted curve).
The growth rate of the field is determined from the slope of the black dashed curves.
Inset: typical $B_z$ field profile during the linear phase.}
\end{figure}

We investigate a simulation with plasma density $n_{p0}=10^{8} \, \rm cm^{-3}$ and photon density $n_{\omega 0} = 10^{22} \, \rm cm^{-3}$.
\textcolor{black}{
We employ the $\delta$ notation for a Dirac delta function, and can write the momentum distribution of the photons as $f=f_{x}=\delta (p_x-100 \,mc)$.}
Figure~\ref{fig:03} reports the energy of the $B_z$ field, denoted by $\mathcal{E}_{B_z}$ and normalized by the incident photon energy $\mathcal{E}_{\omega 0}$ (blue solid curve).
We first focus on the monoenergetic photon distribution (blue solid curve).
The growth rate for all modes $\mathbf{k}$ is $\omega_p^{-1} \Gamma = 7.4 \times 10^{-2}$.
The $B_z$ field profile is displayed during the linear phase of the instability (see the inset) and shows the dominant mode is oblique: $\mathbf{k} = (k_x,k_y) \simeq (1.8, 1.4) \, \omega_p/c$.
Its growth rate is $\omega_p^{-1} \Gamma = 0.12$ close to $\omega_p^{-1} \Gamma \simeq 0.118$, the theoretical prediction from Eq.~\eqref{eq:04}.
\textcolor{black}{The linear stage of the instability starts after a propagation distance $ \simeq 90 \, c/\omega_p $, which is consistent with the minimum inferred theoretically of $34 \, c/\omega_p$ in Eq.~\eqref{eq:r_min}.}
The theoretical upper bound of the energy conversion efficiency given by Eq.~\eqref{eq:06} is $\eta \lesssim 1.5\times 10^{-17}$, which is on the order of the simulation result $\eta = 5\times 10^{-18}$.
We also checked that the pair beam density rises slowly ($+30\%$) on the instability time scale for \textcolor{black}{$\omega_p t = 90 -150$.}
This legitimates a posteriori the use of estimates from linear theory.
\textcolor{black}{However, during the saturation stage, for $\omega_p t = 150 -500$, the front beam density increases from $n_{p0}$ up to $\sim 4 n_{p0}$, a value which is in agreement with Eq.~\eqref{eq:n_beam_sol_sim}.
As a consequence, the beam keeps on filamenting and the magnetic field energy keeps increasing.
}

The aforementioned discussions are focused on a monoenergetic gamma-ray beam.
We intend to extend these results for more complex distributions.
Firstly, we consider a photon distribution $f=f_x f_y$ with $f_x=\delta (p_x-100 \,mc)$ and $f_y$ \textcolor{black}{a Maxwell-J\"{u}ttner distribution of temperature $T_{\gamma,\perp}$.
In the case where $T_{\gamma,\perp}/mc^2 \ll \Delta p_{\perp}/mc$, with $\Delta p_{\perp}/mc=\sqrt{7 \epsilon/(6 \ln \epsilon)}$ given by Eq~\eqref{eq:01}, the results we have discussed before are not changed since the transverse momentum spread of the beam is only due to the Compton scattering kinematics.
In the other limit $T_{\gamma,\perp}/mc^2 \gtrsim \sqrt{7 \epsilon/(6 \ln \epsilon)}$, the transverse momentum spread of the beam is determined by $T_{\gamma,\perp}$.
}
Since the Compton cross section is beamed for gamma-rays, the pair beam presents a similar transverse momentum spread as the photon beam, which is known to lower the growth rate of the OBI~\citep{PREBret2010}.
\textcolor{black}{
The latter behavior is confirmed in Fig.~\ref{fig:03} where we choosed $T_{\gamma,\perp}=3mc^2$.
This implies a transverse momentum spread of $\Delta p_{\perp}/mc = 6$ for the photon source, while the Compton-induced spread is $\Delta p_{\perp}/mc=\sqrt{7 \epsilon/(6 \ln \epsilon)}=5$.
As expected, this leads to a reduction of the growth rate $\omega_p^{-1}\Gamma = 7.4\times 10^{-2} \rightarrow 4.5 \times 10^{-2}$, see the orange dashed curve.}
Secondly, we consider the photons have a monoenergetic distribution $f=f_x$ with $f_x=\delta (p_x-100 \,mc)$ but their density profile has an exponential shape $\propto \exp(-x/100)$ of scale length $100 \, \rm c/ \omega_p$, followed by a flat profile of density $n_{\omega 0}$.
Since the pair beam density is proportional to the photon density, its profile first presents a density gradient and then the triangular shape observed in Figs.~\ref{fig:01}(b3-c3).
As a result, the growth of the OBI is delayed in time, slightly lower but noticeable ($\omega_p^{-1}\Gamma \simeq 1.0\times 10^{-2}$).
This time delay is comparable to the gradient length, as can be seen on the red dotted curve in Fig.~\ref{fig:03}.
Thirdly, we model a synchrotron distribution for the incident photon beam.
We set up this energy distribution as a longitudinal momentum distribution for the photons $f=f_x=F(\eta, \chi)$~\citep{ZETFKlepikov1954}.
\textcolor{black}{
The function $F$ represents the synchrotron spectra of an electron with a quantum parameter $\eta$,  $\chi$ being the photon quantum parameter.}
It has a peak at an energy of $\hbar \omega/mc^2 = 100$.
Photons with an energy of $10$-$100 \, \rm MeV$ are Compton scattered and contribute to increase the pair beam density compared to the monoenergetic case.
Indeed, the Klein-Nishina cross section is a decreasing function of photon energy.
As a result, the growth of the $B_z$ field energy is faster: $\omega_p^{-1}\Gamma = 7.4\times 10^{-2} \rightarrow 1.2 \times 10^{-1}$, as seen on the green dash-dotted curve in Fig.~\ref{fig:03}.

\section{Relevance for astrophysics and laboratory environments \label{sec:05}}

We now discuss under which conditions this process can be observed in astrophysics and in the laboratory.

\begin{figure}
\centering
\includegraphics[width=0.60\textwidth]{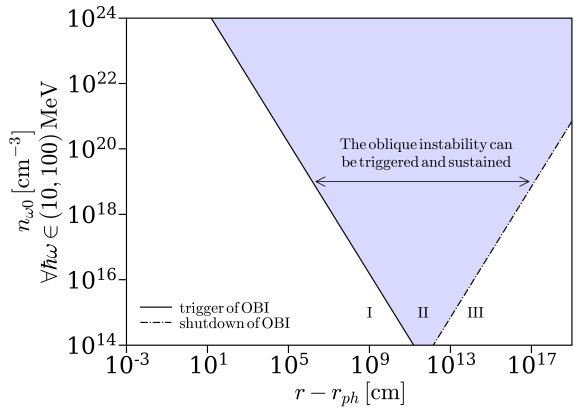}
\caption{
\label{fig:04}
Relevance of the pair beam formation and magnetisation in the frame of the GRB-CBM interaction at the photospheric radius $r_{\rm ph} \sim 10^{12} \, \rm cm$.
\textcolor{black}{
On the left side of the continuous line (area I): one expects only pair beam formation.
Between the continuous and dash-dotted lines (area II), one also awaits the beam-plasma instability.
On the right side of the dash-dotted (area III), the instability is shutdown due to the transverse dilution of the beam.
}
}
\end{figure}

\textcolor{black}{
Figure~\ref{fig:04} covers a large range of photon densities $n_{\omega 0}$ and distances $r-r_{\rm ph}$, with $r_{\rm ph}$ the photospheric radius of a GRB.
It reports whether the propagation distance is large enough to enable the instability to be triggered, see Eq.~\eqref{eq:r_min}.
The latter limit is plotted by the continous line in Fig.~\ref{fig:04} and enables to distinguish two regimes of interaction.
The first one is represented by all the area on the left side of the continuous line and is denoted I.
In this case, the photons are expected to form a relativistic electron positron beam via Compton scattering.
However, the pair beam density is too low and its transverse momentum spread is too high to enable the onset of the instability as detailed in Sec.~\ref{sec:03}.
The second interaction regime is depicted by the area between the continuous and dash-dotted lines and is denoted II.
In this regime, the pair beam density grows high enough to allow the instability to develop.
The dash-dotted line illustrates the maximum length of the filament, as estimated in Eq.~\eqref{eq:r_max}.
It corresponds to the radius above which the beam density is depleted by dilution effects faster than it is loaded via Compton scattering.
}

\textcolor{black}{
For the photospheric radius of a GRB, we evaluated a range of relevant gamma-ray densities.
With estimates from the fireball model, this gamma-ray density (10-100 MeV) is determined to be $n_{\omega}\in(10^{14},10^{22}) \, \rm cm^{-3}$.
From this order of magnitude, we observe in Fig.~\ref{fig:04} that such conditions can trigger the instability and its development on large distances.
To be more precise, the instability can be expected for GRBs whenever their gamma-ray density at the photosphere radius is above \textcolor{black}{$ 10^{14} \, \rm cm^{-3}$}, which corresponds to a GRB with an isotropic equivalent luminosity above $3 \times 10^{50} \, \rm erg/s$.
The second sizeable result is that the beam density remains high enough to sustain the instability on large distances, despite its transverse dilution.
For example, the instability can be maintained to distances $ \geq 10^{17} \, \rm cm $ for gamma-ray densities $ \geq 10^{19} \, \rm cm^{-3}$, and it corresponds to an isotropic equivalent luminosity $\geq 10^{52} \, \rm erg/s$.
}

We now detail the estimates to infer what is the gamma-ray density at the photosphere radius, following~\cite{PRKumar2015}.
Let us consider a compact object with a radius $\sim 10^{7} \, \rm cm$ and an isotropic equivalent luminosity $\mathcal{L}$ in $\mathrm{erg/s}$.
For simplicity we provide the estimate in the compact object frame, thus neglecting the red shift of its host galaxy.
The fireball emerges from the compact object and experiences an adiabatic expansion.
\textcolor{black}{
At the photospheric radius $\sim 10^{12} \, \rm cm$, the emergent thermal radiation from the photosphere has a luminosity which is equivalent to a blackbody temperature of $[1.3 \, \mathrm{MeV}] \times \mathcal{L}_{52}^{1/4}$.
}
The notation $\mathcal{L}_{52}$ is defined as $\mathcal{L}_{52}=\mathcal{L}/(10^{52} \, \rm erg/s)$.
We assume typical GRBs have an isotropic equivalent luminosity in the range $\mathcal{L}\in (10^{47}-10^{54}) \, \rm erg/s$, in line with data recently gathered~\citep{APJAbbott2017}.
This leads to the conclusion that \textcolor{black}{this blackbody equivalent temperature for the luminosity} at the photospheric radius ranges from $70 \, \mathrm{keV}$ up to $ 4 \, \mathrm{MeV}$.
However, the isotropic equivalent luminosity is defined for photons in the standard energy band $1 \, \mathrm{keV}$ up to $ 10 \, \mathrm{MeV}$.
We then deduce the fraction of photons in the energy range $10$-$100 \, \mathrm{MeV}$, where our results are valid, by assuming a black body energy distribution.
\textcolor{black}{
For instance, this ratio of photons (10-100 MeV) becomes $\gtrsim 5\%$ if the GRB isotropic equivalent luminosity is $\gtrsim 10^{52} \, \rm erg/s$.}
The final step is to deduce a gamma-ray density from this total luminosity in gamma-rays.
To this purpose, we assumed the GRB ejecta is a spherically expanding shell and its thickness is the GRB duration, typically ranging between $0.1$ to $100 \, \rm s$.

\textcolor{black}{
We now discuss the role of an external magnetic field that can be expected in an astrophysical scenario.
It is awaited that an external and strong enough magnetic field aligned with the flow will prevent the development of the current filamentation instability~\citep{PRLMolvig1975}.
It was however demonstrated that if the field is not striclty aligned with the flow, then the instability is ensured to take place with a reduced growth rate for a non-relativistic beam temperature~\citep{PoPBret2011}.
This growth rate is a fraction of the growth rate without any magnetic field.
Its detailed value in the low and high magnetisation limits can be found in Eqs.(8) and (10) of the article of~\cite{PoPBret2011}.
}

It is worth to question whether this Compton-driven beam formation and magnetisation can take place in the laboratory.
Despite the ongoing worldwide efforts, no large size and confined pair plasmas have been formed in a laboratory environment yet.
One can still mention that pair jets of high density $n_{p0}\sim 10^{16}\, \rm cm^{-3}$ can be generated by fast electrons going through a mm-sized high-Z targets. The fast electrons can be generated by an intense laser~\citep{PoPChen2015,SRLiang2015}, or by a plasma wakefield accelerator~\citep{NCSarri2015,PoPXu2016}.
\textcolor{black}{Recent numerical work bring forward a new scheme to create dense pair jets $n_{p0}\sim 10^{13}\, \rm cm^{-3}$ with size of several skin-depths in all directions from the interaction of relativistic protons ($400 \, \rm GeV/c$) with a solid beryllium/lead target~\citep{ArxivArrowsmith2020}.}
Regarding gamma-ray sources (10-100 MeV), we estimated a density $n_{\omega 0}\sim 10^{17} \, \rm cm^{-3}$ for the experimental photon source recently obtained by Non Linear Inverse Compton scattering~\citep{PRXPoder2018,PRXCole2018}.
\textcolor{black}{For such photon density, one would need a plasma length of $10^8 \, \rm cm$ to witness the instability, which cannot be achieved in a laboratory environment}.
We also consider the results of a recent simulation from an extremely dense gamma-ray source obtained during the interaction of a dense electron beam with multiple micron-sized foils~\citep{PRLSampath2021}.
Although it remains a numerical result and its experimental realisation may lie far in the future, it is achieving a record gamma-ray density of \textcolor{black}{$n_{\omega 0}\sim 10^{23} \, \rm cm^{-3}$} with a distribution peaked at 100 MeV.
One can see in Fig.~\ref{fig:04} that even this high gamma-ray density might be enough to witness the onset of the Compton-driven instability in a laboratory frame.
\textcolor{black}{One would need a plasma length of $150 \, \rm cm$ to observe it.
}
Overall, our results indicate that exploration of the process of beam formation and magnetisation in the laboratory is unlikely in the short term due to the combined requirements of plasma length and gamma-ray densities.

\section{Conclusions \label{sec:06}}

To summarize, we reported a study of gamma-ray propagation in a background pair plasma.
We showed that it can lead to the formation of a relativistic pair beam, thanks to the beaming of the Compton cross section for photons above 10 MeV.
\textcolor{black}{A theoretical model to describe the formation of the beam driven by Compton scattering has been developed, and compared with PIC simulations.}
We showed the pair beam can achieve a relativistic Lorentz factor with a density comparable to the background plasma.
In addition, we quantified the transverse momentum spread of the beam, which is induced by the Compton cross section.
We demonstrated that the pair beam, as it propagates, can convert its kinetic energy to magnetic energy via the oblique instability, although limited by its transverse momentum spread.
The conversion efficiency of this process is low\textcolor{black}{, but it can occur over long distances}.
We extrapolated from PIC simulations that it could lead to the generation of magnetic fields on distances larger than a parsec (up to $\simeq 10^{25} \, \rm cm$) for a perfectly collimated GRB ejecta.
\textcolor{black}{For a less collimated one ($\theta \leq 20$), we estimated the instability can be sustained over long distances $10^{12}\rightarrow 10^{17} \, \rm cm$ for GRBs with an isotropic equivalent luminosity ranging from $3\times 10^{50} \rightarrow 10^{52} \, \rm erg/s$.}
We showed that these results can be used to address the energy dissipation of gamma-rays at the photospheric radius of GRBs.
We checked the robustness of our results for various types of photon sources, with non-uniform density profiles and transverse and longitudinal spreads in their momentum distribution.
\textcolor{black}{The simulations done in the frame of this study are performed} for the typical orders of magnitude for plasma and photon densities expected at the photospheric radius of a GRB, using estimates from the hot fireball model as well as data recently gathered on GRBs.

This work offers two directions to explore in the future.
The first could be to consider photon distributions with energies below 10 MeV and connect with previous works on Compton driven plasma wakes~\citep{APJFrederiksen2008,PRLGaudio2020}.
The second is to understand the impact of the beam formation and magnetization on global GRB models.

\section*{Acknowledgments}

The authors acknowledge fruitful discussions with Dr. Del Gaudio and Dr. Schoeffler.
This work was supported by the European Research Council (ERC-2015-AdG Grant 695088).
We also acknowledge PRACE for awarding access to MareNostrum based in Spain.

\bibliographystyle{jpp}

\begin{thebibliography}{48}
  \expandafter\ifx\csname natexlab\endcsname\relax\def\natexlab#1{#1}\fi
  \def\au#1{#1} \def\ed#1{#1} \def\yr#1{#1}\def\at#1{#1}\def\jt#1{\textit{#1}}
    \def\bt#1{#1}\def\bvol#1{\textbf{#1}} \def\vol#1{#1} \def\pg#1{#1}
    \def\publ#1{#1}\def\arxiv#1{#1}\def\org#1{#1}\def\st#1{\textit{#1}}
  
  \bibitem[Abbott {\em et~al.\/}(2017)Abbott, Abbott, Abbott, Acernese, Ackley,
    Adams, Adams, Addesso, Adhikari, Adya \& Collaboration]{APJAbbott2017}
  {\sc \au{Abbott, B.~P.}, \au{Abbott, R.}, \au{Abbott, T.~D.}, \au{Acernese,
    F.}, \au{Ackley, K.}, \au{Adams, C.}, \au{Adams, T.}, \au{Addesso, P.},
    \au{Adhikari, R.~X.}, \au{Adya, V.~B.} \& \au{Collaboration, LIGO~VIRGO}}
    \yr{2017}  \at{Gravitational waves and gamma-rays from a binary neutron star
    merger: {GW}170817 and {GRB} 170817a}.  \jt{The Astrophysical Journal}
    \bvol{848}~(2),  \pg{L13}.
  
  \bibitem[Acciari {\em et~al.\/}(2019)Acciari, Ansoldi, Antonelli, Engels,
    Baack, Babi{\'c}, Banerjee, Barres~de Almeida, Barrio, Gonz{\'a}lez \& {MAGIC
    Collaboration}]{NatAcciari2019}
  {\sc \au{Acciari, V.~A.}, \au{Ansoldi, S.}, \au{Antonelli, L.~A.}, \au{Engels,
    A.~Arbet}, \au{Baack, D.}, \au{Babi{\'c}, A.}, \au{Banerjee, B.},
    \au{Barres~de Almeida, U.}, \au{Barrio, J.~A.}, \au{Gonz{\'a}lez, J.~Becerra}
    \& \au{{MAGIC Collaboration}}} \yr{2019}  \at{Observation of inverse compton
    emission from a long $\gamma$-ray burst}.  \jt{Nature}  \bvol{575}~(7783),
    \pg{459--463}.
  
  \bibitem[Arrowsmith {\em et~al.\/}(2020)Arrowsmith, Shukla, Charitonidis, Boni,
    Chen, Davenne, Froula, Huffman, Kadi, Reville, Richardson, Sarkar, Shaw,
    Silva, Trines, Bingham \& Gregori]{ArxivArrowsmith2020}
  {\sc \au{Arrowsmith, C.~D.}, \au{Shukla, N.}, \au{Charitonidis, N.}, \au{Boni,
    R.}, \au{Chen, H.}, \au{Davenne, T.}, \au{Froula, D.~H.}, \au{Huffman,
    B.~T.}, \au{Kadi, Y.}, \au{Reville, B.}, \au{Richardson, S.}, \au{Sarkar,
    S.}, \au{Shaw, J.~L.}, \au{Silva, L.~O.}, \au{Trines, R. M. G.~M.},
    \au{Bingham, R.} \& \au{Gregori, G.}} \yr{2020} Generating ultra-dense pair
    beams using 400 gev/c protons,  \arxiv{arXiv: 2011.04398}.
  
  \bibitem[{Beloborodov}(2005)]{APJBeloborodov2005}
  {\sc \au{{Beloborodov}, A.~M.}} \yr{2005}  \at{{Afterglow Emission from
    Pair-loaded Blast Waves in Gamma-Ray Bursts}}.  \jt{The Astrophysical
    Journal}  \bvol{627}~(1),  \pg{346--367}.
  
  \bibitem[Blumenthal \& Gould(1970)]{RMPBlumenthal1970}
  {\sc \au{Blumenthal, G.~R.} \& \au{Gould, R.~J.}} \yr{1970}
    \at{Bremsstrahlung, synchrotron radiation, and compton scattering of high
    energy electrons traversing dilute gases}.  \jt{Review of Modern Physics}
    \bvol{42},  \pg{237--270}.
  
  \bibitem[Bret \& Alvaro(2011)]{PoPBret2011}
  {\sc \au{Bret, A.} \& \au{Alvaro, E.~Perez}} \yr{2011}  \at{Robustness of the
    filamentation instability as shock mediator in arbitrarily oriented magnetic
    field}.  \jt{Physics of Plasmas}  \bvol{18}~(8),  \pg{080706}.
  
  \bibitem[Bret {\em et~al.\/}(2005{\natexlab{{\em a\/}}})Bret, Firpo \&
    Deutsch]{PRLBret2005}
  {\sc \au{Bret, A.}, \au{Firpo, M.-C.} \& \au{Deutsch, C.}}
    \yr{2005{\natexlab{{\em a\/}}}}  \at{Characterization of the initial
    filamentation of a relativistic electron beam passing through a plasma}.
    \jt{Physical Review Letters}  \bvol{94},  \pg{115002}.
  
  \bibitem[Bret {\em et~al.\/}(2005{\natexlab{{\em b\/}}})Bret, Firpo \&
    Deutsch]{PREBret2005}
  {\sc \au{Bret, A.}, \au{Firpo, M.-C.} \& \au{Deutsch, C.}}
    \yr{2005{\natexlab{{\em b\/}}}}  \at{Electromagnetic instabilities for
    relativistic beam-plasma interaction in whole $k$ space: Nonrelativistic beam
    and plasma temperature effects}.  \jt{Phys. Rev. E}  \bvol{72},  \pg{016403}.
  
  \bibitem[Bret {\em et~al.\/}(2010)Bret, Gremillet \& B\'enisti]{PREBret2010}
  {\sc \au{Bret, A.}, \au{Gremillet, L.} \& \au{B\'enisti, D.}} \yr{2010}
    \at{Exact relativistic kinetic theory of the full unstable spectrum of an
    electron-beam--plasma system with {M}axwell-{J}\"uttner distribution
    functions}.  \jt{Physical Review E}  \bvol{81},  \pg{036402}.
  
  \bibitem[Cavallo \& Rees(1978)]{MNRASCavallo1978}
  {\sc \au{Cavallo, G.} \& \au{Rees, M.~J.}} \yr{1978}  \at{{A qualitative study
    of cosmic fireballs and gamma-ray bursts}}.  \jt{Monthly Notices of the Royal
    Astronomical Society}  \bvol{183}~(3),  \pg{359--365}.
  
  \bibitem[Chen {\em et~al.\/}(2015)Chen, Link, Sentoku, Audebert, Fiuza, Hazi,
    Heeter, Hill, Hobbs, Kemp, Kemp, Kerr, Meyerhofer, Myatt, Nagel, Park,
    Tommasini \& Williams]{PoPChen2015}
  {\sc \au{Chen, H.}, \au{Link, A.}, \au{Sentoku, Y.}, \au{Audebert, P.},
    \au{Fiuza, F.}, \au{Hazi, A.}, \au{Heeter, R.~F.}, \au{Hill, M.}, \au{Hobbs,
    L.}, \au{Kemp, A.~J.}, \au{Kemp, G.~E.}, \au{Kerr, S.}, \au{Meyerhofer,
    D.~D.}, \au{Myatt, J.}, \au{Nagel, S.~R.}, \au{Park, J.}, \au{Tommasini, R.}
    \& \au{Williams, G.~J.}} \yr{2015}  \at{The scaling of electron and positron
    generation in intense laser-solid interactions}.  \jt{Physics of Plasmas}
    \bvol{22}~(5),  \pg{056705}.
  
  \bibitem[Cole {\em et~al.\/}(2018)Cole, Behm, Gerstmayr, Blackburn, Wood,
    Baird, Duff, Harvey, Ilderton, Joglekar, Krushelnick, Kuschel, Marklund,
    McKenna, Murphy, Poder, Ridgers, Samarin, Sarri, Symes, Thomas, Warwick,
    Zepf, Najmudin \& Mangles]{PRXCole2018}
  {\sc \au{Cole, J.~M.}, \au{Behm, K.~T.}, \au{Gerstmayr, E.}, \au{Blackburn,
    T.~G.}, \au{Wood, J.~C.}, \au{Baird, C.~D.}, \au{Duff, M.~J.}, \au{Harvey,
    C.}, \au{Ilderton, A.}, \au{Joglekar, A.~S.}, \au{Krushelnick, K.},
    \au{Kuschel, S.}, \au{Marklund, M.}, \au{McKenna, P.}, \au{Murphy, C.~D.},
    \au{Poder, K.}, \au{Ridgers, C.~P.}, \au{Samarin, G.~M.}, \au{Sarri, G.},
    \au{Symes, D.~R.}, \au{Thomas, A. G.~R.}, \au{Warwick, J.}, \au{Zepf, M.},
    \au{Najmudin, Z.} \& \au{Mangles, S. P.~D.}} \yr{2018}  \at{{Experimental
    Evidence of Radiation Reaction in the Collision of a High-Intensity Laser
    Pulse with a Laser-Wakefield Accelerated Electron Beam}}.  \jt{Phys. Rev. X}
    \bvol{8},  \pg{011020}.
  
  \bibitem[Del~Gaudio {\em et~al.\/}(2020{\natexlab{{\em a\/}}})Del~Gaudio,
    Fonseca, Silva \& Grismayer]{PRLGaudio2020}
  {\sc \au{Del~Gaudio, F.}, \au{Fonseca, R.~A.}, \au{Silva, L.~O.} \&
    \au{Grismayer, T.}} \yr{2020{\natexlab{{\em a\/}}}}  \at{Plasma wakes driven
    by photon bursts via compton scattering}.  \jt{Phys. Rev. Lett.}  \bvol{125},
     \pg{265001}.
  
  \bibitem[Del~Gaudio {\em et~al.\/}(2020{\natexlab{{\em b\/}}})Del~Gaudio,
    Grismayer, Fonseca \& Silva]{JPPGaudio2020}
  {\sc \au{Del~Gaudio, F.}, \au{Grismayer, T.}, \au{Fonseca, R.~A.} \& \au{Silva,
    L.~O.}} \yr{2020{\natexlab{{\em b\/}}}}  \at{Compton scattering in
    particle-in-cell codes}.  \jt{Journal of Plasma Physics}  \bvol{86}~(5),
    \pg{905860516}.
  
  \bibitem[Fonseca {\em et~al.\/}(2002)Fonseca, Silva, Tsung, Decyk, Lu, Ren,
    Mori, Deng, Lee, Katsouleas \& Adam]{PrFonseca2002}
  {\sc \au{Fonseca, R.~A.}, \au{Silva, L.~O.}, \au{Tsung, F.~S.}, \au{Decyk,
    V.~K.}, \au{Lu, W.}, \au{Ren, C.}, \au{Mori, W.~B.}, \au{Deng, S.}, \au{Lee,
    S.}, \au{Katsouleas, T.} \& \au{Adam, J.~C.}} \yr{2002} Osiris: A
    three-dimensional, fully relativistic particle in cell code for modeling
    plasma based accelerators.  \bt{In {\em Computational Science --- ICCS
    2002\/} (ed. \ed{Peter M.~A. Sloot, Alfons~G. Hoekstra, C.~J.~Kenneth Tan \&
    Jack~J. Dongarra})},  \pg{pp. 342--351}.  \publ{Berlin, Heidelberg: Springer
    Berlin Heidelberg}.
  
  \bibitem[Frederiksen(2008)]{APJFrederiksen2008}
  {\sc \au{Frederiksen, J.~T.}} \yr{2008}  \at{Stochastically induced gamma-ray
    burst wakefield processes}.  \jt{The Astrophysical Journal}  \bvol{680}~(1),
    \pg{L5--L8}.
  
  \bibitem[Gruzinov \& M{\'{e}}sz{\'{a}}ros(2000)]{APJGruzinov2000}
  {\sc \au{Gruzinov, A.} \& \au{M{\'{e}}sz{\'{a}}ros, P.}} \yr{2000}  \at{Photon
    acceleration in variable ultrarelativistic outflows and high-energy spectra
    of gamma-ray bursts}.  \jt{The Astrophysical Journal}  \bvol{539}~(1),
    \pg{L21--L24}.
  
  \bibitem[Haugb{\o}lle {\em et~al.\/}(2013)Haugb{\o}lle, Frederiksen \&
    Nordlund]{PoPHaugbolle2013}
  {\sc \au{Haugb{\o}lle, T.}, \au{Frederiksen, J.~T.} \& \au{Nordlund, {\AA}.}}
    \yr{2013}  \at{photon-plasma: A modern high-order particle-in-cell code}.
    \jt{Physics of Plasmas}  \bvol{20}~(6),  \pg{062904}.
  
  \bibitem[Klein \& Nishina(1928)]{NatKlein1928}
  {\sc \au{Klein, O.} \& \au{Nishina, Y.}} \yr{1928}  \at{The scattering of light
    by free electrons according to {D}irac's new relativistic dynamics}.
    \jt{Nature}  \bvol{122}~(3072),  \pg{398--399}.
  
  \bibitem[Klepikov(1954)]{ZETFKlepikov1954}
  {\sc \au{Klepikov, N.~P.}} \yr{1954}  \at{Emission of photons or
    electron-positron pairs in magnetic fields}.  \jt{Journal of Experimental and
    Theoretical Physics}  \bvol{26},  \pg{19}.
  
  \bibitem[Kumar \& Zhang(2015)]{PRKumar2015}
  {\sc \au{Kumar, P.} \& \au{Zhang, B.}} \yr{2015}  \at{The physics of gamma-ray
    bursts and relativistic jets}.  \jt{Physics Reports}  \bvol{561},  \pg{1 --
    109}.
  
  \bibitem[{Levinson} \& {Cerutti}(2018)]{AALevinson2018}
  {\sc \au{{Levinson}, A.} \& \au{{Cerutti}, B.}} \yr{2018}
    \at{{Particle-in-cell simulations of pair discharges in a starved
    magnetosphere of a Kerr black hole}}.  \jt{Astronomy {\&} Astrophysics}
    \bvol{616},  \pg{A184}.
  
  \bibitem[{Liang} {\em et~al.\/}(2015){Liang}, {Clarke}, {Henderson}, {Fu},
    {Lo}, {Taylor}, {Chaguine}, {Zhou}, {Hua}, {Cen}, {Wang}, {Kao}, {Hasson},
    {Dyer}, {Serratto}, {Riley}, {Donovan} \& {Ditmire}]{SRLiang2015}
  {\sc \au{{Liang}, E.}, \au{{Clarke}, T.}, \au{{Henderson}, A.}, \au{{Fu}, W.},
    \au{{Lo}, W.}, \au{{Taylor}, D.}, \au{{Chaguine}, P.}, \au{{Zhou}, S.},
    \au{{Hua}, Y.}, \au{{Cen}, X.}, \au{{Wang}, X.}, \au{{Kao}, J.},
    \au{{Hasson}, H.}, \au{{Dyer}, G.}, \au{{Serratto}, K.}, \au{{Riley}, N.},
    \au{{Donovan}, M.} \& \au{{Ditmire}, T.}} \yr{2015}  \at{{High e+/e- Ratio
    Dense Pair Creation with 10$^{21}$W.cm$^{-2}$ Laser Irradiating Solid
    Targets}}.  \jt{Scientific Reports}  \bvol{5},  \pg{13968}.
  
  \bibitem[{Lightman}(1982)]{APJLightman1982}
  {\sc \au{{Lightman}, A.~P.}} \yr{1982}  \at{{Relativistic thermal plasmas -
    Pair processes and equilibria}}.  \jt{The Astrophysical Journal}  \bvol{253},
     \pg{842--858}.
  
  \bibitem[Lyutikov(2006)]{NJPLyutikov2006}
  {\sc \au{Lyutikov, M}} \yr{2006}  \at{The electromagnetic model of gamma-ray
    bursts}.  \jt{New Journal of Physics}  \bvol{8}~(7),  \pg{119--119}.
  
  \bibitem[Madau \& Thompson(2000)]{APJMadau2000a}
  {\sc \au{Madau, P.} \& \au{Thompson, C.}} \yr{2000}  \at{Relativistic winds
    from compact gamma-ray sources. i. radiative acceleration in the
    klein-nishina regime}.  \jt{The Astrophysical Journal}  \bvol{534}~(1),
    \pg{239--247}.
  
  \bibitem[Martins {\em et~al.\/}(2009)Martins, Fonseca, Silva \&
    Mori]{APJMartins2009}
  {\sc \au{Martins, S.~F.}, \au{Fonseca, R.~A.}, \au{Silva, L.~O.} \& \au{Mori,
    W.~B.}} \yr{2009}  \at{{Ion} {Dynamics} {and} {Acceleration} {in}
    {Relativistic} {Shocks}}.  \jt{The Astrophysical Journal}  \bvol{695}~(2),
    \pg{L189--L193}.
  
  \bibitem[Medvedev \& Loeb(1999)]{APJMedvedev1999}
  {\sc \au{Medvedev, M.~V.} \& \au{Loeb, A.}} \yr{1999}  \at{Generation of
    magnetic fields in the relativistic shock of gamma-ray burst sources}.
    \jt{The Astrophysical Journal}  \bvol{526}~(2),  \pg{697--706}.
  
  \bibitem[Mehlhaff {\em et~al.\/}(2020)Mehlhaff, Werner, Uzdensky \&
    Begelman]{MNRASMehlhaff2020}
  {\sc \au{Mehlhaff, J~M}, \au{Werner, G~R}, \au{Uzdensky, D~A} \& \au{Begelman,
    M~C}} \yr{2020}  \at{{Kinetic beaming in radiative relativistic magnetic
    reconnection: a mechanism for rapid gamma-ray flares in jets}}.  \jt{Monthly
    Notices of the Royal Astronomical Society}  \bvol{498}~(1),  \pg{799--820}.
  
  \bibitem[{Meszaros} \& {Rees}(1993)]{APJMeszaros1993}
  {\sc \au{{Meszaros}, P.} \& \au{{Rees}, M.~J.}} \yr{1993}  \at{{Relativistic
    Fireballs and Their Impact on External Matter: Models for Cosmological
    Gamma-Ray Bursts}}.  \jt{The Astrophysical Journal}  \bvol{405},  \pg{278}.
  
  \bibitem[Meszaros \& Rees(1997)]{APJMeszaros1997}
  {\sc \au{Meszaros, P.} \& \au{Rees, M.~J.}} \yr{1997}  \at{Optical and
    long‐wavelength afterglow from gamma‐ray bursts}.  \jt{The Astrophysical
    Journal}  \bvol{476}~(1),  \pg{232–237}.
  
  \bibitem[Molvig(1975)]{PRLMolvig1975}
  {\sc \au{Molvig, K.}} \yr{1975}  \at{Filamentary instability of a relativistic
    electron beam}.  \jt{Phys. Rev. Lett.}  \bvol{35},  \pg{1504--1507}.
  
  \bibitem[{Narayan} {\em et~al.\/}(1992){Narayan}, {Paczynski} \&
    {Piran}]{APJNarayan1992}
  {\sc \au{{Narayan}, R.}, \au{{Paczynski}, B.} \& \au{{Piran}, T.}} \yr{1992}
    \at{{Gamma-Ray Bursts as the Death Throes of Massive Binary Stars}}.  \jt{The
    Astrophysical Journal}  \bvol{395},  \pg{L83}.
  
  \bibitem[Piran(2005)]{RMPPiran2005}
  {\sc \au{Piran, T.}} \yr{2005}  \at{The physics of gamma-ray bursts}.
    \jt{Review of Modern Physics}  \bvol{76},  \pg{1143--1210}.
  
  \bibitem[Poder {\em et~al.\/}(2018)Poder, Tamburini, Sarri, Di~Piazza, Kuschel,
    Baird, Behm, Bohlen, Cole, Corvan, Duff, Gerstmayr, Keitel, Krushelnick,
    Mangles, McKenna, Murphy, Najmudin, Ridgers, Samarin, Symes, Thomas, Warwick
    \& Zepf]{PRXPoder2018}
  {\sc \au{Poder, K.}, \au{Tamburini, M.}, \au{Sarri, G.}, \au{Di~Piazza, A.},
    \au{Kuschel, S.}, \au{Baird, C.~D.}, \au{Behm, K.}, \au{Bohlen, S.},
    \au{Cole, J.~M.}, \au{Corvan, D.~J.}, \au{Duff, M.}, \au{Gerstmayr, E.},
    \au{Keitel, C.~H.}, \au{Krushelnick, K.}, \au{Mangles, S. P.~D.},
    \au{McKenna, P.}, \au{Murphy, C.~D.}, \au{Najmudin, Z.}, \au{Ridgers, C.~P.},
    \au{Samarin, G.~M.}, \au{Symes, D.~R.}, \au{Thomas, A. G.~R.}, \au{Warwick,
    J.} \& \au{Zepf, M.}} \yr{2018}  \at{Experimental signatures of the quantum
    nature of radiation reaction in the field of an ultraintense laser}.
    \jt{Phys. Rev. X}  \bvol{8},  \pg{031004}.
  
  \bibitem[Racusin {\em et~al.\/}(2009)Racusin, Liang, Burrows, Falcone,
    Sakamoto, Zhang, Zhang, Evans \& Osborne]{APJRacusin2009}
  {\sc \au{Racusin, J.~L.}, \au{Liang, E.~W.}, \au{Burrows, D.~N.}, \au{Falcone,
    A.}, \au{Sakamoto, T.}, \au{Zhang, B.~B.}, \au{Zhang, B.}, \au{Evans, P.} \&
    \au{Osborne, J.}} \yr{2009}  \at{Jet breaks and energetics of {S}wift
    gamma-ray burst x-ray afterglows}.  \jt{The Astrophysical Journal}
    \bvol{698}~(1),  \pg{43--74}.
  
  \bibitem[Rees \& Meszaros(1994)]{APJRees1994}
  {\sc \au{Rees, M.~J.} \& \au{Meszaros, P.}} \yr{1994}  \at{Unsteady outflow
    models for cosmological gamma-ray bursts}.  \jt{The Astrophysical Journal}
    \bvol{430},  \pg{L93}.
  
  \bibitem[Sampath {\em et~al.\/}(2021)Sampath, Davoine, Corde, Gremillet,
    Gilljohann, Sangal, Keitel, Ariniello, Cary, Ekerfelt, Emma, Fiuza, Fujii,
    Hogan, Joshi, Knetsch, Kononenko, Lee, Litos, Marsh, Nie, O'Shea, Peterson,
    Claveria, Storey, Wu, Xu, Zhang \& Tamburini]{PRLSampath2021}
  {\sc \au{Sampath, A.}, \au{Davoine, X.}, \au{Corde, S.}, \au{Gremillet, L.},
    \au{Gilljohann, M.}, \au{Sangal, M.}, \au{Keitel, C.~H.}, \au{Ariniello, R.},
    \au{Cary, J.}, \au{Ekerfelt, H.}, \au{Emma, C.}, \au{Fiuza, F.}, \au{Fujii,
    H.}, \au{Hogan, M.}, \au{Joshi, C.}, \au{Knetsch, A.}, \au{Kononenko, O.},
    \au{Lee, V.}, \au{Litos, M.}, \au{Marsh, K.}, \au{Nie, Z.}, \au{O'Shea, B.},
    \au{Peterson, J.~R.}, \au{Claveria, P. S.~M.}, \au{Storey, D.}, \au{Wu, Y.},
    \au{Xu, X.}, \au{Zhang, C.} \& \au{Tamburini, M.}} \yr{2021}  \at{Extremely
    dense gamma-ray pulses in electron beam-multifoil collisions}.  \jt{Physicsl
    Review Letters}  \bvol{126},  \pg{064801}.
  
  \bibitem[{Sarri} {\em et~al.\/}(2015){Sarri}, {Poder}, {Cole}, {Schumaker}, {Di
    Piazza}, {Reville}, {Doria}, {Dromey}, {Gizzi}, {Green}, {Grittani}, {Kar},
    {Keitel}, {Krushelnick}, {Kushel}, {Mangles}, {Najmudin}, {Thomas}, {Vargas}
    \& {Zepf}]{NCSarri2015}
  {\sc \au{{Sarri}, G.}, \au{{Poder}, K.}, \au{{Cole}, J.}, \au{{Schumaker}, W.},
    \au{{Di Piazza}, A.}, \au{{Reville}, B.}, \au{{Doria}, D.}, \au{{Dromey},
    B.}, \au{{Gizzi}, L.}, \au{{Green}, A.}, \au{{Grittani}, G.}, \au{{Kar}, S.},
    \au{{Keitel}, C.~H.}, \au{{Krushelnick}, K.}, \au{{Kushel}, S.},
    \au{{Mangles}, S.}, \au{{Najmudin}, Z.}, \au{{Thomas}, A.~G.~R.},
    \au{{Vargas}, M.} \& \au{{Zepf}, M.}} \yr{2015}  \at{{Generation of a
    neutral, high-density electron-positron plasma in the laboratory}}.
    \jt{Nature Communications}  \bvol{6}.
  
  \bibitem[Silva {\em et~al.\/}(2003)Silva, Fonseca, Tonge, Dawson, Mori \&
    Medvedev]{APJSilva2003}
  {\sc \au{Silva, L.~O.}, \au{Fonseca, R.~A.}, \au{Tonge, J.~W.}, \au{Dawson,
    J.~M.}, \au{Mori, W.~B.} \& \au{Medvedev, M.~V.}} \yr{2003}
    \at{Interpenetrating plasma shells: Near-equipartition magnetic field
    generation and nonthermal particle acceleration}.  \jt{The Astrophysical
    Journal}  \bvol{596}~(1),  \pg{L121--L124}.
  
  \bibitem[Silva {\em et~al.\/}(2002)Silva, Fonseca, Tonge, Mori \&
    Dawson]{PoPSilva2002}
  {\sc \au{Silva, L.~O.}, \au{Fonseca, R.~A.}, \au{Tonge, J.~W.}, \au{Mori,
    W.~B.} \& \au{Dawson, J.~M.}} \yr{2002}  \at{On the role of the purely
    transverse {W}eibel instability in fast ignitor scenarios}.  \jt{Physics of
    Plasmas}  \bvol{9}~(6),  \pg{2458--2461}.
  
  \bibitem[Spitkovsky(2008)]{APJSpitkovsky2008}
  {\sc \au{Spitkovsky, A.}} \yr{2008}  \at{Particle acceleration in relativistic
    collisionless shocks: Fermi process at last?}  \jt{The Astrophysical Journal}
     \bvol{682}~(1),  \pg{L5--L8}.
  
  \bibitem[Stern(2003)]{MNRASStern2003}
  {\sc \au{Stern, B.~E.}} \yr{2003}  \at{{Electromagnetic catastrophe in
    ultrarelativistic shocks and the prompt emission of gamma-ray bursts}}.
    \jt{Monthly Notices of the Royal Astronomical Society}  \bvol{345}~(2),
    \pg{590--600}.
  
  \bibitem[Thompson \& Madau(2000)]{APJThompson2000}
  {\sc \au{Thompson, C.} \& \au{Madau, P.}} \yr{2000}  \at{Relativistic winds
    from compact gamma-ray sources. {II}. pair loading and radiative acceleration
    in gamma-ray bursts}.  \jt{The Astrophysical Journal}  \bvol{538}~(1),
    \pg{105--114}.
  
  \bibitem[Uzdensky(2011)]{SSRUzdensky2011}
  {\sc \au{Uzdensky, D.~A.}} \yr{2011}  \at{Magnetic reconnection in extreme
    astrophysical environments}.  \jt{Space Science Reviews}  \bvol{160}~(1),
    \pg{45--71}.
  
  \bibitem[Xu {\em et~al.\/}(2016)Xu, Shen, Xu, Li, Yu, Li, Lu, Wang, Wang,
    Liang, Leng, Li \& Xu]{PoPXu2016}
  {\sc \au{Xu, T.}, \au{Shen, B.}, \au{Xu, J.}, \au{Li, S.}, \au{Yu, Y.}, \au{Li,
    J.}, \au{Lu, X.}, \au{Wang, C.}, \au{Wang, X.}, \au{Liang, X.}, \au{Leng,
    Y.}, \au{Li, R.} \& \au{Xu, Z.}} \yr{2016}  \at{Ultrashort megaelectronvolt
    positron beam generation based on laser-accelerated electrons}.  \jt{Physics
    of Plasmas}  \bvol{23}~(3),  \pg{033109}.
  
  \bibitem[Yang {\em et~al.\/}(1994)Yang, Arons \& Langdon]{PoPYang1994}
  {\sc \au{Yang, T.‐Y.~Brian}, \au{Arons, Jonathan} \& \au{Langdon, A.~Bruce}}
    \yr{1994}  \at{Evolution of the {W}eibel instability in relativistically hot
    electron–positron plasmas}.  \jt{Physics of Plasmas}  \bvol{1}~(9),
    \pg{3059--3077}.
  
  \bibitem[Zhang \& Yan(2010)]{APJZhang2010}
  {\sc \au{Zhang, B.} \& \au{Yan, H.}} \yr{2010}  \at{The internal collision
    induced magnetic reconnection and turbulence ({ICMART}) model of gamma-ray
    bursts}.  \jt{The Astrophysical Journal}  \bvol{726}~(2),  \pg{90}.
  
  \end{thebibliography}

\end{document}